\documentclass[aps,twocolumn,showpacs,floatfix,superscriptaddress]{revtex4-1}
\usepackage{graphicx}
\usepackage{psfrag}
\begin{document}
\newcommand{\beq}{\begin{equation}}
\newcommand{\eeq}{\end{equation}}
\newcommand{\ben}{\begin{eqnarray}}
\newcommand{\een}{\end{eqnarray}}
\newcommand{\bea}{\begin{array}}
\newcommand{\eea}{\end{array}}
\newcommand{\om}{(\omega )}
\newcommand{\bef}{\begin{figure}}
\newcommand{\eef}{\end{figure}}
\newcommand{\leg}[1]{\caption{\protect\rm{\protect\footnotesize{#1}}}}
\newcommand{\ew}[1]{\langle{#1}\rangle}
\newcommand{\be}[1]{\mid\!{#1}\!\mid}
\newcommand{\no}{\nonumber}
\newcommand{\etal}{{\em et~al }}
\newcommand{\geff}{g_{\mbox{\it{\scriptsize{eff}}}}}
\newcommand{\da}[1]{{#1}^\dagger}
\newcommand{\cf}{{\it cf.\/}\ }
\newcommand{\ie}{{\it i.e.\/}\ }
\title{Do your volatility smiles take care of extreme events?}
\author{L.~Spadafora}  
\affiliation {Dipartimento di Matematica e Fisica,
Universit\`a Cattolica, via Musei 41, 25121 Brescia, Italy}
\author{G.~P.~Berman}
\affiliation {Theoretical Division, MS-B213, Los Alamos National Laboratory, Los Alamos, NM, 87545}
\author{F.~Borgonovi}
\affiliation {Dipartimento di Matematica e Fisica,
Universit\`a Cattolica, via Musei 41, 25121 Brescia, Italy}
\affiliation{I.N.F.N. Sezione di Pavia, Pavia, Italy}
\begin{abstract}

In the Black-Scholes context we consider the probability distribution function (PDF) of financial returns implied by volatility smile and we study the relation between the decay of its tails and the fitting parameters of the smile. We show that, considering a scaling law derived from data, it is possible to get a new fitting procedure of the volatility smile that considers also the exponential decay of the real PDF of returns observed in the financial markets. Our study finds application in the Risk Management activities where the tails characterization of financial returns PDF has a central role for the risk estimation. 
\end{abstract}
\date{\today}
\maketitle

%%%%%%%%%%%%%%%%%%%%%%%%%%%%

\section{Introduction}

%%%%%%%%%%%%%%%%%%%%%%%%%%%%
Financial derivatives are the modern financial instruments that are used in many activities and for different purposes: mitigating risk exposure, speculation and arbitrage, trading strategies, providing leverage, etc. Knowing the fair value of such kind of contracts is not, generally, an easy task and it is of crucial importance, for example,  for the correct evaluation of a portfolio of financial instruments and the related risks. \\
One of the simplest  ``products" on the derivative financial market is the European
call (put) option~\cite{Hull, Wilmott}.
Considering the risk neutral approach, the price of the European call option, $C \equiv  C(S_T, K, T, r)$,
is defined by
\begin{equation}\label{pricing_risk_neutral}
C = e^{-rT} \int_{K}^{\infty} (S_T - K) P(S_T) d S_T,
\end{equation}
where $S_T$ is the stock price at time $t = T$, $K$ is the strike price of the option, $T$ is the 
expiration time (time to maturity) of the option, $r$ 
is the interest rate and $P(S_T) \ge 0$ is the distribution function of the stock prices in a ``risk-neutral world" ($\int_{0}^{\infty} P(S_T) d S_T =1$).

Eq.~(\ref{pricing_risk_neutral}) is too general because it does not make any hypothesis on the underlying stock price distribution function, $P(S_T)$. To calculate explicitly the option price, $C$, using Eq.~(\ref{pricing_risk_neutral}), one can assume that the distribution function, $P(S_T)$, is log-normal, so that, for the logarithmic return deprived of the risk-free 
component, $x = \ln(S_T/S(t)) - r(T-t)$, the distribution is normal:

\begin{equation}\label{normal_dist}
P(x) = \frac{1}{\sqrt{2\pi \sigma^2 (T - t)}} 
\exp\left[-\frac{(x + \sigma^2(T-t)/2)^2}{2 \sigma^2 (T-t)} \right],
\end{equation}
where $S(t)$  is the stock price at time $t$ and $\sigma$ is the stock price volatility. For seek of simplicity, in the following we consider $t = 0$ and we define $S_0 \equiv S(t=0)$.

Using the Eqs.~(\ref{pricing_risk_neutral}),~(\ref{normal_dist}) it is possible to get an explicit expression 
for the price of the European call option:

\begin{equation}\label{BS_solution}
C^{BS} = S_0 N(d_1) - K e^{-rT}N(d_2),
\end{equation}
where
\begin{equation}
\begin{array}{ccc}
d_1 &=& \displaystyle\frac{\ln(S_0/K)+ (r + \sigma^2/2)T}{\sigma \sqrt{T}},  \\
&&\\
d_2 & =& d_1 - \sigma \sqrt{T}, \\
&&\\
N(x) & = \displaystyle & \frac{1}{\sqrt{2\pi}}\int_{-\infty}^{x} \ dz \ e^{-{z^2}/{2}}.\\
&&\\
\end{array}
\end{equation}

The Eq.~(\ref{BS_solution}) gives an analytical solutions for the European call option pricing and it is the main results of the Black-Scholes (BS) theory about option pricing~\cite{Black}. 

The distribution function (\ref{normal_dist}) follows from a stochastic model for stock prices,
\begin{equation}\label{log_norm}
dS = r S dt + \sigma S dz,
\end{equation}
where $dz$ is a Wiener increment~\cite{winer_note}.
It can be shown it is never optimal to exercise an American call 
option on a non-dividend-paying stock early~\cite{Hull,Bouchaud}; 
therefore Eq.~(\ref{BS_solution}) can also be used to estimate the fair value for this kind of options. \\
Since 1973, when the BS model was published Quantitative Finance became a prominent aspect in many banks and financial institutions activities and a lot of new, more realistic models were developed for the option pricing~\cite{Heston, Hagan}. 
%%% Citare Heston, SABR...
These new models are currently implemented and used by traders and risk managers of many financial institutions and it could seem that the BS model is by now outdated and irrelevant for financial applications. On the contrary, because of its simplicity and the small number of parameters, BS model is still a benchmark, used by practitioners in many situations where getting a reliable calibration of parameters of more complex models could be unattainable in practice. Simplicity and a sort of reluctance to changes explain, in our opinion, the reason why, after about 37 years from its publication, BS model is still used by practitioners and justify the importance of our study to get a correct calibration procedure for the volatility smile (VS) effect also from a theoretical point of view. In a recent paper~\cite{Spada} it is shown a new calibration procedure that can be obtained using an adiabatic approach to avoid arbitrage opportunities. The term ``adiabatic'' comes from comes from statistical physics and is related to the slowness of the variation of a parameter $\lambda$ that specifies the properties of a system or an external field. In fact from a physical point of view, it can be shown that  if in a system one introduces a small perturbation ($\lambda$) compared to  the characteristics period of the motion $T$, namely:
\begin{equation}
T \frac{d\lambda}{d t} \ll \lambda
\end{equation}
the rate of the change of the energy of the system will be also small~\cite{Landau}. In the same spirit we assume that our parameter $\lambda$ is represented by the implied volatility $\sigma$ and we study how to characterize PDF of returns with a small perturbation of the parameter $\sigma$ to get a suitable description of actual data, coherent from a theoretical point of view. 
In particular, in the following, it is shown the importance of this calibration procedure from the risk management point of view and its relevance in the risk estimation.\\
The rest of the paper is organized as follows.\\

In Section \ref{VS_real}, we analyze the volatility smiles from Foreign Exchange (FX) market data and we propose a suitable function to fit it. We also find a relation between the fitting parameters that holds for every symmetric smile that help us to identify the real independent variables of our system. Using this relation we determine a suitable range of parameters for our simulation.

In Section \ref{risk} we outline the relation between VS and PDF of returns and we stress the importance of getting a suitable fit for the VS for the risk estimation

In Section \ref{mu_decay}, we study the relation between the parameters of our volatility smile function and the decay of the PDF of financial returns and we find an equation to describe this kind of behavior.

In Section \ref{recipe}, we show how to get a more reliable fit of the volatility smile, considering the exponential decay of the returns PDF. To do this we exploit the relation between the standard deviation of an exponential distribution and its decay. 
Finally, in Section \ref{conclusions} we present our conclusions.

%%%%%%%%%%%%%%%%%%%%%%%%%%%%%%%%%%%%%%%%%%%%%%%%%%%%%%%%%%%%%%%%%%%%%%%%%%%%%%%%%%%%%%

\section{Volatility smile: Analysis of actual market data}\label{VS_real}

%%%%%%%%%%%%%%%%%%%%%%%%%%%%%%%%%%%%%%%%%%%%%%%%%%%%%%%%%%%%%%%%%%%%%%%%%%%%%%%%%%%%%%

Typically, traders on option markets and practitioners consider the BS model
 as a zeroth order approximation that takes into 
account the main features of options prices. To get a pricing closer to the actual data, they consider the volatility as a
parameter that can be adjusted considering the inverse problem given by Eq.~(\ref{BS_solution}) and the
 real price of
call and put options. In this way  a more reliable value of the 
volatility (\emph{implied volatility}) can be obtained and
it can be used to price more complex options for which analytical solutions are not available. 
The value of the implied volatility depends on the value of the strike, $K$, in a well-known 
characteristic curve called  the \emph{smile volatility} 
(typically for foreign currency options) whose shape is approximately parabolic and 
symmetric, or \emph{skew volatility}
(typically for equity options) when asymmetric effects dominate~\cite{Risk, Tompkins,
Toft, Campa}.

An intuitive explanation of this shape can be found if an actual returns distribution is considered.
In fact, it is well known that the tails of the returns PDF are not Gaussian but exhibit a 
power law decay (fat-tails)~\cite{Sorn, Stan} or exponential decay~\cite{Heston}.
%% Citare PDF di heston
On the contrary,  BS model assumes that the PDF of returns is Gaussian thus underestimating the actual probability of rare events. To compensate for this model deficiency, one has to consider the greater implied volatility for strike out of the money than for strike  at the money. 
%%% Inserire ref HULL 

In this paper we focus our attention on the VS of foreign currency options and we neglect the skew effect~\cite{Kirch}. 
To perform our analysis we consider the volatility smile as a function of the $\Delta$ of the 
option (defined by Eq.~({\ref{delta})), the time to maturity, $T$, and  the currency 
considered. We consider specific days, for which volatility is not affected by the skew effect, and we use Bloomberg as data provider.
%%%% Inserire giorni considerati
In the BS model, the $\Delta$  of a call options is defined as:
\begin{equation}
\Delta = \frac{\partial C}{\partial S_T} = \displaystyle erf(d_1),
\label{delta}
\end{equation}
Inverting this relation is possible to get an expression for $x$:
\begin{equation}\label{Delta_to_x}
x = \sigma^2/2 T - \sigma \sqrt{T} \ erf^{-1} (\Delta),
\end{equation}
where $erf^{-1}(x)$ is the inverse of the error function.
%%%% Inserire figura smile e fit relativo
%%%% Ricorda di cambiare i numeri perch� qui l'unit� temporale � il giorno...
\begin{figure}[t]
\begin{center}
\includegraphics[scale=0.2]{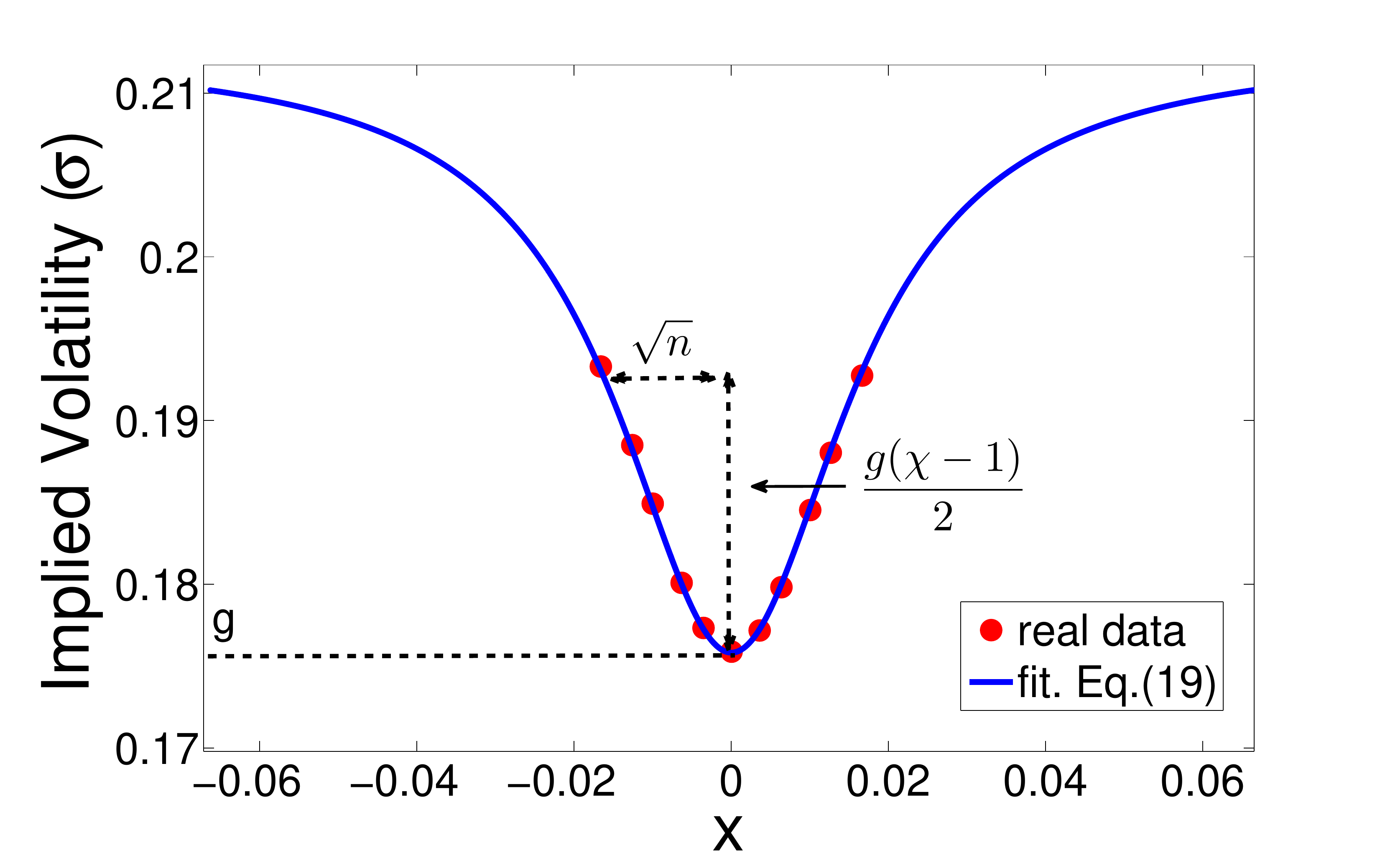}
\end{center}
\caption{Typical VS and the 
relative fit obtained with Eq.~ (\ref{sigma_fit}). 
The parameters of the fit are: $g= 0.1758(5) $, $\chi=1.20(9)
$, $n=0.00030(9) $.
 We get the real data using Bloomberg provider and they refer to the AUDUSD currency with
time to maturity $T=1/365$ years.} 
\label{figfitsmile}
\end{figure}
In Fig.~\ref{figfitsmile}
we show an example of VS in terms of our variables and a 
suitable fit given by the function:
\begin{equation}\label{sigma_fit}
\sigma(x) = g\left[ 1 + (\chi-1)  \frac{  (x + g^2T/2)^2}{(x + g^2T/2)^2 + n}\right]
\end{equation}
where $g, \chi, n$ are fitting parameters. In this case, $g$ represents the minimum of the volatility smile, $\sqrt{n}$ is the half width at the half height, while 
$g(\chi - 1)$ represents
the height of the smile. In particular $\chi$ is the ratio 
between the limiting value of $\sigma$ as $x$ approaches $\infty$ and $g$ .
In this way the variation of $\sigma$ is bounded between $g$ and $g\chi$.
In the light of the intuitive explanation of the volatility smile proposed above 
and since from~(\ref{normal_dist}) it follows that the average value, 
\begin{equation}
\langle x \rangle  = -\frac{\sigma^2 T}{2},
\end{equation}
one expects that the minimum of the implied volatility occurs at $ x = -g^2T/2 $ 
as required by our fitting function. 

We repeat the fitting procedure considering the  volatility smile
for different days, currencies and time to maturity $T$ (Table~\ref{dataset_VS}),
 then we analyze the relations between the fitting parameters.  

\begin{table}[htbp]
\begin{center}
\begin{tabular}{l c c}
\hline
\bfseries{Currency} & \bfseries{Maturities} (days)& \bfseries{Date} \\
\hline
AUDUSD, EURCHF & $1, 7, 14, 21, 30$ & $21/10/2009$ \\
\hline
 EURGBP, EURJPY& $60 ,90, 120, 180, 270$ & $01/02/2010$ \\
\hline
EURUSD, GBPUSD         & $360, 540, 720, 1080$ & $01/04/2010$ \\
\hline
USDCAD, USDCHF & & \\
\hline
\end{tabular}
\end{center}
\caption{Dataset for VS}
\label{dataset_VS}
\end{table} 

As already observed in \cite{Spada}, the following
relation between $n, T, g$ holds:
\begin{equation}\label{scaling_1}
\sqrt{n} = c g \sqrt{T}
\end{equation}
where $c = 2.65(28)$ is a fitting parameter. 
Our intuitive explanation of this equation is really simple and it is related to the fact 
that the PDF of returns is not Gaussian but exhibits fat/exponential tails. 
Indeed, while the term $\sqrt{n}$  gives the order of magnitude 
 of the   volatility amplitude, 
$g\sqrt{T}$ represents the minimum of the implied volatility
(which can be considered as the unperturbed standard deviation of the PDF of returns).
Therefore  Eq.~(\ref{scaling_1}) suggests
that when $x$ is about 
$2-3$ times the standard deviation of the returns distribution (namely in the tails) the 
implied volatility should be increased to fatten up the PDF of returns.\\
In Section \ref{mu_decay} we use this relation to fix the typical range of 
parameters of the VS in order to perform  suitable simulations for the description
 of actual data.

%%%%%%%%%%%%%%%%%%%%%%%%%%%%%%%%%%%%%%%%%%%%%%%%%%%%%%%%%%%%%%%%%%%%%%%%%%%%%%%%%%%%%%%%%%%%%%%%%%%%%%%

\section{Importance of VS in Risk Estimation}\label{risk}

%%%%%%%%%%%%%%%%%%%%%%%%%%%%%%%%%%%%%%%%%%%%%%%%%%%%%%%%%%%%%%%%%%%%%%%%%%%%%%%%%%%%%%%%%%%%%%%%%%%%%%%
In Eq.~(\ref{pricing_risk_neutral}) the distribution function, $P(S_T)$, can be rather 
arbitrary but it is natural to assume that $P(S_T)$ does not depend on the strike price, $K$. According to Eq.~(\ref{pricing_risk_neutral}), the option price, $C$, is expressed explicitly through the strike price, $K$.
Differentiating $C$ in Eq.~(\ref{pricing_risk_neutral}) twice with respect to $K$, we have~\cite{Malz},

\begin{equation}\label{price_distr}
P(S_T ) = e^{rT} \frac{\partial^2 C(K)}{\partial K^2} \biggr|_{K=S_T}.
\end{equation}
In Eq.~(\ref{price_distr}), we indicate only the dependence $C(K)$ in the option prices. Eq.~(\ref{price_distr}) makes explicit the relation between a pricing model, given by $C(K)$, and the implicit distribution of prices (and, by a simple change of variables, of financial returns), assuming a risk neutral approach. For example, if one consider the BS model for call options pricing (Eq.~(\ref{BS_solution})), using Eq.~(\ref{price_distr}) one gets, as expected, a Gaussian distribution for financial returns. More generally, if one considers the dependence, $\sigma = \sigma(K)$, in Eq.~(\ref{BS_solution}), it is possible to get the analytical expression of the implied distribution of financial returns~\cite{Spada}:

\begin{equation}\label{distr_ret_pert}
P_\sigma (x) = 
\frac{1}{\sqrt{2 \pi \sigma^2 T} } \exp\left[ - 
\frac{( x  + x_0)^2}{2\sigma^2 T} \right] F(x;  T,\sigma),
\end{equation}
where we have defined:
\begin{equation}\label{perturb_fact}
\begin{array}{ccc}
&& x \equiv \ln\left (\frac{K}{S_0}\right)\biggr|_{K = S_T} - r T,\\
&&\\
&& F(x; T,\sigma) = \displaystyle
(1 - \frac{\sigma^\prime}{\sigma} x  )^2 - \frac{(\sigma^\prime\sigma T)^2}{4} +  
 \sigma \sigma^{\prime \prime} T,\\
&&\\
&& \sigma^\prime = \displaystyle  \frac{\partial \sigma}{\partial x}, \\
&&\\
&& \sigma^{\prime \prime}  =  \displaystyle \frac{\partial^2 \sigma}{\partial x^2}. 
\end{array}
\end{equation}
From (\ref{perturb_fact}) it is clear that if $\sigma$ is constant, Eq.~(\ref{distr_ret_pert}) gives the Gaussian distribution for the standard Black-Scholes
 model.\\
It is also helpful to define the implied complementary cumulative distribution function (CCDF) of financial returns as:
\begin{equation}
E(x)  =  1 - \int_{-\infty}^x P_{\sigma}(y) dy.
\end{equation}
Eq.~(\ref{distr_ret_pert}) shows that there is a strong relation between $VS$ and
the  PDF of financial returns. From another point of view, Eq.~(\ref{distr_ret_pert})
 should be seen as a warning that shows how similar fits of a VS could
 imply strong differences in the implied returns of the 
PDF with obvious consequences, for example, on the risk estimation. If one considers, for example, the two curves (red and blue) in Fig.~(\ref{smile_comp}), it is clear that even if the two lines are close to the actual data, the differences in the decay of the two distributions can be relevant with important consequences for the risk estimation procedure. 
%%% Figura di confronto smile
\begin{figure}[t]
\begin{center}
\includegraphics[scale=0.2]{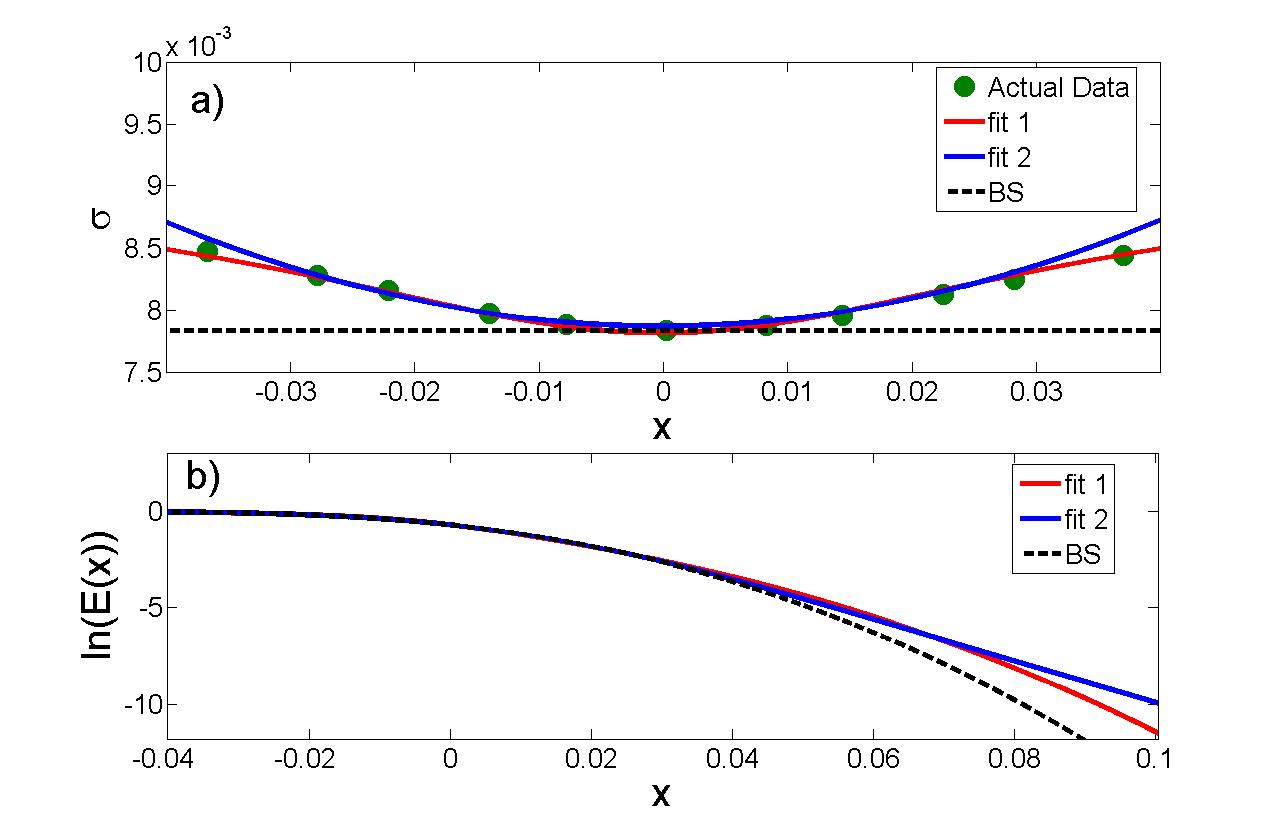}
\end{center}
\caption{Comparison of two suitable approximations for the VS (red and blue) (a) and their CCDFs (b). As evident, even if the two curves can be close to the actual data, the differences in the Value at Risk estimation can be relevant. For comparison we also show the case of a completely flat smile (black) and its Gaussian distribution.} 
\label{smile_comp} 
\end{figure}  

One could consider, for example, the estimation of the risk using the standard
 VAR (value-at-risk) measure~\cite{Bouchaud}, defined as

\begin{equation}
\mathcal{P_{VAR}} = \int_{-\infty}^{-\Lambda_{VAR}} P(x) dx,
\end{equation} 
where $\Lambda_{VAR}$ represents our estimation of the maximum potential loss with a 
fixed confidence level given by  $\mathcal{P_{VAR}}$ and $P(x)$ is a generic function that represents reuturns PDF. In this paper,
 we consider $P(x) = P_{\sigma}(x)$ and $\mathcal{P_{VAR}} = 1\%$ as a standard value for the confidence level; 
this means we can expect a loss less than or equal to
 $\Lambda_{VAR}$ in the $99\%$ of the cases.\\
For the distributions in Fig.~(\ref{smile_comp}), we get $\Lambda_{VAR}^{red} = 5.23\%$ and $\Lambda_{VAR}^{blue} = 5.06\%$, so the difference in the VAR estimation using the two different fits is about $3.27\%$. To have an idea of the order of magnitude of the error, one should consider that for the flat smile (BS) in the figure, we get $\Lambda_{VAR}^{BS} = 4.8\%$ and the difference with the other VAR estimation is about $5\%-8\%$. \\
From this example it is clear there is some arbitrariness
 in the fitting parameters of the VS function that can generate significant differences in the description of the implied returns distribution, with important consequences, for example, from the risk estimation point of view. So that the importance of getting a reliable fitting procedure consistent with the theoretical aspects, as already stressed in \cite{Ciliberti}.\\
In this framework, we focus our attention on the generalized BS model by considering VS effect and we try to characterize the decay of the tails of the implied distribution of returns as a function of the fitting parameters of the VS, to get a suitable procedure for the smile fitting coherent with the historical observed decay of the actual returns PDFs. As already shown, a suitable characterization of the implied distributions decay can have a fundamental importance, for example, for the risk estimation. 

%%%%%%%%%%%%%%%%%%%%%%%%%%%%%%%%%%%%%%%%%%%%%%%%%%%%%%%%%%%%%%%%%%%%%%%%%%%%%%%
\section{Relation between VS and the tails of PDF of financial returns}\label{mu_decay}
%%%%%%%%%%%%%%%%%%%%%%%%%%%%%%%%%%%%%%%%%%%%%%%%%%%%%%%%%%%%%%%%%%%%%%%%%%%%%%%

In this Section, we want to establish a simple relation between the parameters of the fitting function Eq.~(\ref{sigma_fit}) and the decay of the tails of the implied distribution of returns, Eq.~(\ref{distr_ret_pert}).
To better understand what we mean for ``decay of the tails", we need to analyze the
 structure of the Eqs.(\ref{sigma_fit}, \ref{distr_ret_pert}). 
First of all, it is important to notice that $\sigma(x)$
is a bounded function
$$ g \leq \sigma(x) \leq g\chi.$$ The whole process can be seen 
as a continuous 
transition from the a minimum value $g$ to a limit value $g\chi$ reached for 
large enough returns, $x$. From the PDF point of view, we can think of the VS as a 
continuous transition between two Gaussian distributions with different standard 
deviations, $g$ and $g\chi$. 
So, due to  our choice of the VS fitting function, we already know that for large 
$x$ values the tails of the
 implied distribution behaves as a Gaussian distribution. 
Nonetheless, there is a region of $x$, namely the \emph{region of the transition}, 
not described by a Gaussian, since in this case $\sigma$ is not constant. 
In Section \ref{VS_real} we have already discussed the
  order of magnitude of $x$ for this region: $x \sim \sqrt{n} = 2.6 g \sqrt{T}$ 
which corresponds to the tail of the distribution. So, even if we know that for really large $x$ the implied distribution is a Gaussian, the region that can be related to the tails of actual returns distributions is the region of transition and this is the region we are going to study in details. \\
Looking at a typical implied distribution of returns on a semilog plot it seems reasonable to approximate the region of the transition by a straight line, as shown in Fig.~(\ref{trans_region}). 

%% Inserire figura con PDF implicita e retta che approssima la regione di trasizione
\begin{figure}[t]
\begin{center}
\includegraphics[scale=0.2]{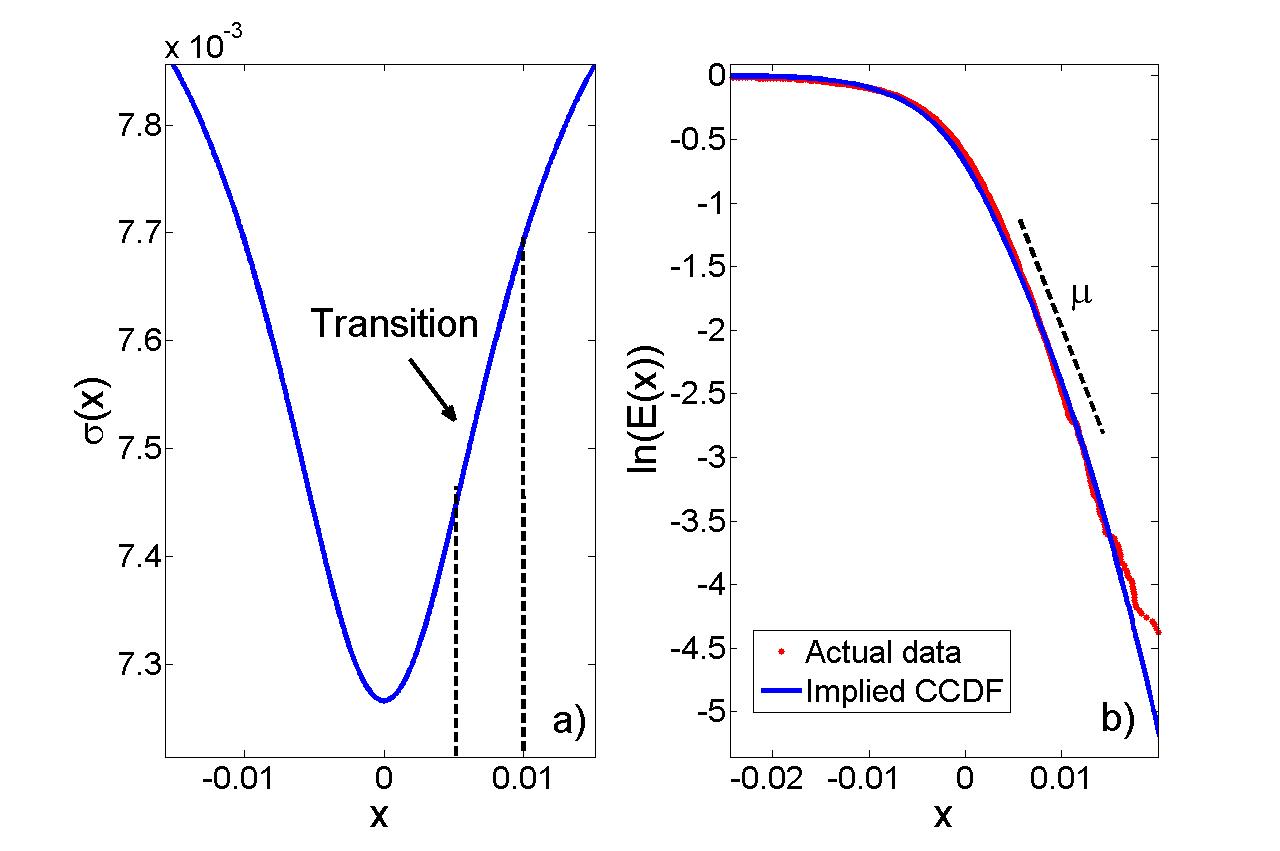}
\end{center}
\caption{a) We show the transition region of the VS $\cal{R} $ and b) the exponential decay approximation for the CCDF of returns in the same region.} 
\label{trans_region} 
\end{figure}  

This approximation is equivalent to assume that the tails distributions of financial returns have an exponential decay, $\exp{(-\mu |x|)}$, where $\mu$ is the factor that characterize the tail. This fact finds confirmation in our real data analysis and it is coherent with results shown in~\cite{Dragulescu}.\\
The main goal of this Section is to establish a relation between the parameter of decay, $\mu$, and the fitting parameters of the smile, $g, \chi, n$. The procedure we consider is straightforward and it is described in the following.\\
First of all, we fixed the range of the parameters repeating many times the fitting procedure and considering the dataset described in Section~\ref{VS_real}. In Table~\ref{range_param}, we show the range of the parameters that we used to perform our simulations (we used the parameter $\rho = n/(g^2T)$ instead of $n$ due to  the scaling relation Eq.~(\ref{scaling_1})).\\
%%%%% Inserire tabella con range dei parametri %%%%%
\begin{table}[htbp]

\begin{center}
\begin{tabular}{l c c}
\hline
 & \bfseries{min} & \bfseries{max} \\
\hline
\bfseries{g} & $0.03$ & $0.5$ \\
\hline
\bfseries{$\rho$} & $2.5$ & $10$ \\
\hline
\bfseries{T} (days) & $1$ & $1080$ \\
\hline
\bfseries{$\chi$} & $1.01$ & $3$ \\
\hline
\end{tabular}
\end{center}
\caption{Range of the parameters of the numerical simulations.}
\label{range_param}
\end{table}

Using this range of parameters, we consider the implied CCDF of returns, derived from Eq.~(\ref{distr_ret_pert}), and we fit the region of transition considering an exponential decay, $\exp{(-\mu |x|)}$, where $\mu$ is the fitting parameter. In this way we get for every set of the parameters in the Table~\ref{range_param} the corresponding decay parameter, $\mu$. 
We define the region of transition  
as ${\cal{R}}  = \{ x | \sqrt{n}/2 \leq x \leq \sqrt{n} \}$; 
in this way, if $A = g\chi + g$ represents the height of the VS, we are considering 
the region from the $20\%$ to the $50\%$ of the total height. \\
Our goal is to find a relation between $\mu$ and the three parameters of the VS. 
First of all, let us fix  $\chi = 1$, so that the VS is completely flat. 
In this case we know that the distribution is Gaussian, $F(x, T, \sigma ) =1 $
and  the parameter $\mu$ should be thought of 
 an approximation of an exponential decay. In this case, 
$\mu$ can be easily estimated as:

\begin{equation}
\mu = \frac{\Delta y}{\Delta x} = \frac{ \ln(\mathcal{P}(\sqrt{n}) - \ln(\mathcal{P}(\sqrt{n}/2))}{\sqrt{n}/2},
\end{equation}
where $\mathcal{P}$ is the CCDF of $P$ defined in Eq.~(\ref{distr_ret_pert}). Performing some calculations we get:

\begin{equation}
\mu_1 = \frac{2}{g\sqrt{T}} f(\rho),
\end{equation}
where, the function $f(\rho)$, is defined by,
\begin{equation}
f(\rho)  =  \frac{1}{\sqrt{\rho}} \ln \left[ 
\frac{ 1 - \mbox{erf}(\frac{1}{2}\sqrt{\frac{\rho}{2}})}{1 -
\mbox{erf}(\sqrt{\frac{\rho}{2}})}\right],
\end{equation}
and has the following asymptotic expansion:
\begin{equation}
f(\rho) \simeq \left\{
\begin{array}{rl}
\sqrt{\rho} & \mbox{if } \rho  \longmapsto +\infty \\
1/2\sqrt{\pi}          & \mbox{if }  \rho  \longmapsto 0.
\end{array}
\right.
\end{equation}
Let us now  discuss the case $\chi\ne 1 $: 
in the light of the adiabatic interpretation presented in \cite{Spada}, 
we  expect that on increasing $\chi$, the PDF 
will present, soon or later a minimum. This means that  the PDF should be flatter than
before, so that $\mu$ should decrease. This is coherent with our physical interpretation of the VS as a small perturbation of a theoretical system represented by a gaussian distribution. Increasing the order of magnitude of the perturbation, here represented by the parameter $\chi$, we get a PDF of returns increasingly different from the gaussian until the adiabatic limit of the perturbation is violated. After that point the system cannot be described by a perturbative approach.

For simplicity, let us assume  the simple inverse proportionality: 

\begin{equation}\label{mu_estim}
\mu_\chi  = \displaystyle\frac{\mu_1}{\chi} =   \frac{2}{\chi g \sqrt{T}} f(\rho).
\end{equation}
Relation (\ref{mu_estim}) has been checked in Fig.~(\ref{mu_model})
where  we plot the real parameter $\mu$ 
obtained by our simulation {\it vs} the parameter $\mu_\chi$ given by (\ref{mu_estim}):
the agreement is within a  $2\%$ of mean squared error.

%% Inserire figura bontà modello per mu
\begin{figure}[t]
\begin{center}
\includegraphics[scale=0.2]{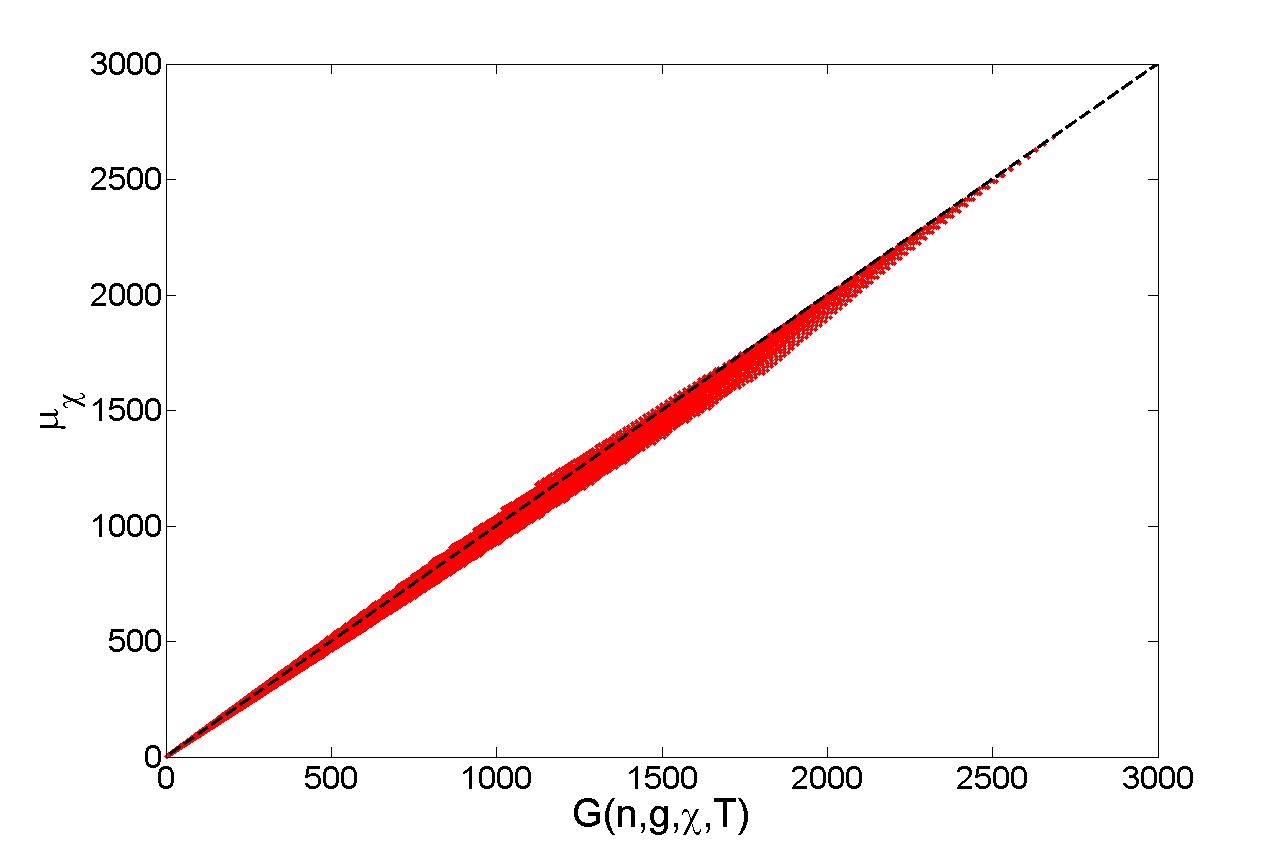}
\end{center}
\caption{We show the relation between the decay parameter
$\mu$, given by numerical simulation and the estimation given by 
Eq.~(\ref{mu_estim}). As reference, we also show the (dotted) line 
$\mu = \mu_\chi$.
 } 
\label{mu_model} 
\end{figure}

%%%%%%%%%%%%%%%%%%%%%%%%%%%%%%%%%%%%%%%%%%%%%%%%%%%%%%%%%%%%%%%%%%%%%%%%%%%%%%%

\section{A new recipe to fit the volatility smile}\label{recipe}

%%%%%%%%%%%%%%%%%%%%%%%%%%%%%%%%%%%%%%%%%%%%%%%%%%%%%%%%%%%%%%%%%%%%%%%%%%%%%%%
In this Section we show how to include the information given by the formula~(\ref{mu_estim}) on the decay of the CCDF of the financial returns to get a suitable fit of the VS coherent from theoretical point of view.  Firstly, to do this we need to analyze what is the ordinary interpretation of the implied volatility of the BS model and its relation with historical volatility. Implied volatility is usually interpreted as the future volatility of the market and represents the traders and practitioners 
vision. From this point of view historical volatility can be interpreted as a peculiar
 realization of this vision at some particular time period. So, in general, 
there will be a mismatch between historical volatility and implied volatility and 
this fact is reflected on historical and implied PDF of returns. Therefore, to use properly
 the information on the decay of the historical distribution, we need at first the 
scaling relation between the volatility and the decay of the distribution. 
This relation can be estimated from historical series of currencies (AUDUSD, EURCHF, 
EURGBP, EURJPY, EURUSD, GBPUSD, USDCAD, USDCHF, time period 2001-2010) using the 
following procedure. We consider different time lag ($T=1, 10, 100 $ days) and build 
different historical series of returns. We divide each series into subgroups of at least 300 elements and we evaluated the standard deviation of each group. To evaluate the decay we consider the CCDF of returns using the procedure described in~\cite{Bouchaud} and we fit the tail decay using a straight line in a semi log plot. We repeat this procedure for any subgroup and for any currency to make explicit the relation between $\mu_H$ and $\sigma_H$.
%% inserire figura relazione mu  vs sigma
\begin{figure}[t]
\begin{center}
\includegraphics[scale=0.2]{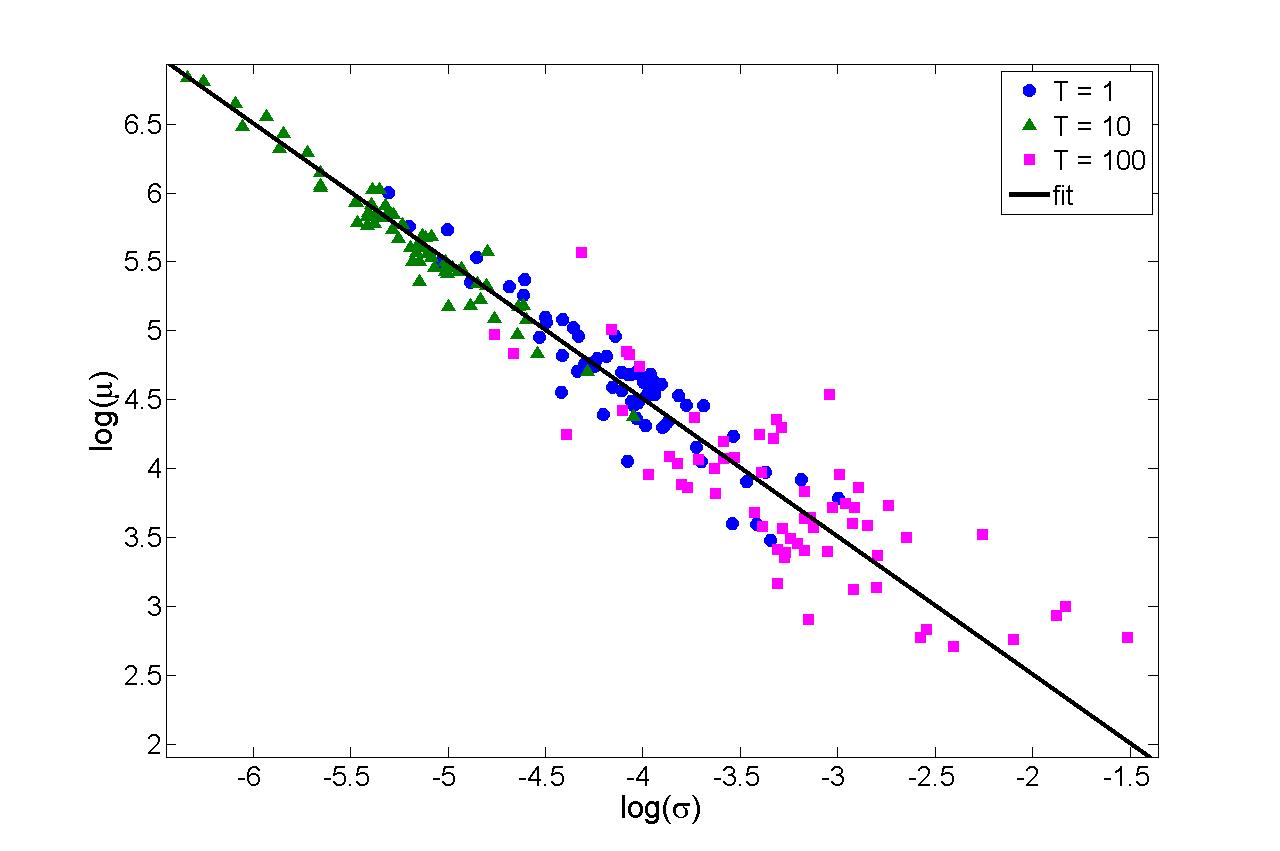}
\end{center}
\caption{Relation (independent on the time lag $T$) 
between $\mu_H$ and $\sigma_H$
 considering three different time lag for the returns 
($T=1, 10, 100$). We also show 
the best linear fit $\ln(\mu_H) =  \ln(\sigma_H) + ln(C_1)$, where 
$C_1 = 1.6\pm 0.5$.}
\label{mu_sigma} 
\end{figure}  
In Fig.~(\ref{mu_sigma}) we show our results 
superimposed with a suitable fitting function
\begin{equation}\label{mu_scaling}
\sigma_H  = \frac{C_1}{\mu_H},
\end{equation}
where $C_1 = 1.6\pm 0.5$ is a fitting parameter.
Let us observe that this is in quite good agreement with an exponential PDF for returns,
since in that case one would have $\sigma_H = 2/\mu_H$.

Eq.~(\ref{mu_scaling}) makes explicit the relation between $\mu_H$ and $\sigma_H$ 
(their product should be a constant $\approx 1.6$) 
and gives us the opportunity to exploit the information on the
 historical decay of the PDF of financial returns to get a suitable fit of the
 VS. The procedure can be summarized as follow:
\begin{itemize}
\item Using the historical price series we determine the decay and the standard 
deviation of the financial returns, respectively: $\mu_H, \sigma_H$.
\item Identifying the product  $\mu_H \sigma_H$  with $g\sqrt{T} \mu_\chi $ and using
our  estimation, Eq.~(\ref{mu_estim}), we can obtain 
one of three fitting parameters, e.g. $\chi$,  describing the VS, as a function of the 
other two ($g, n$) :
\begin{equation}
\chi =  \frac{2}{\mu_H \sigma_H } f(n/g^2T).
\end{equation}

\end{itemize}

Following this approach, we reduce the number of free parameters
 for the smile fitting, fixing implicitly the right decay of the PDF of returns. As already stressed in Section~\ref{risk}, the need of getting a suitable fit for the VS coherent also with the theoretical aspects of the model, is really important in many Risk Management activities and could lead to significant differences in risk estimation.

For example in Fig.~(\ref{appl2}), we compare the PDF of returns obtained by a standard fitting procedure of VS (unconditional fit) with the one obtained following the procedure described before (conditional fit). As evident, even if the two fitting procedures give similar curve for the VS, the effect on the VAR estimation are of the order of $10\%$.

\begin{figure}[t]
\begin{center}
\includegraphics[scale=0.2]{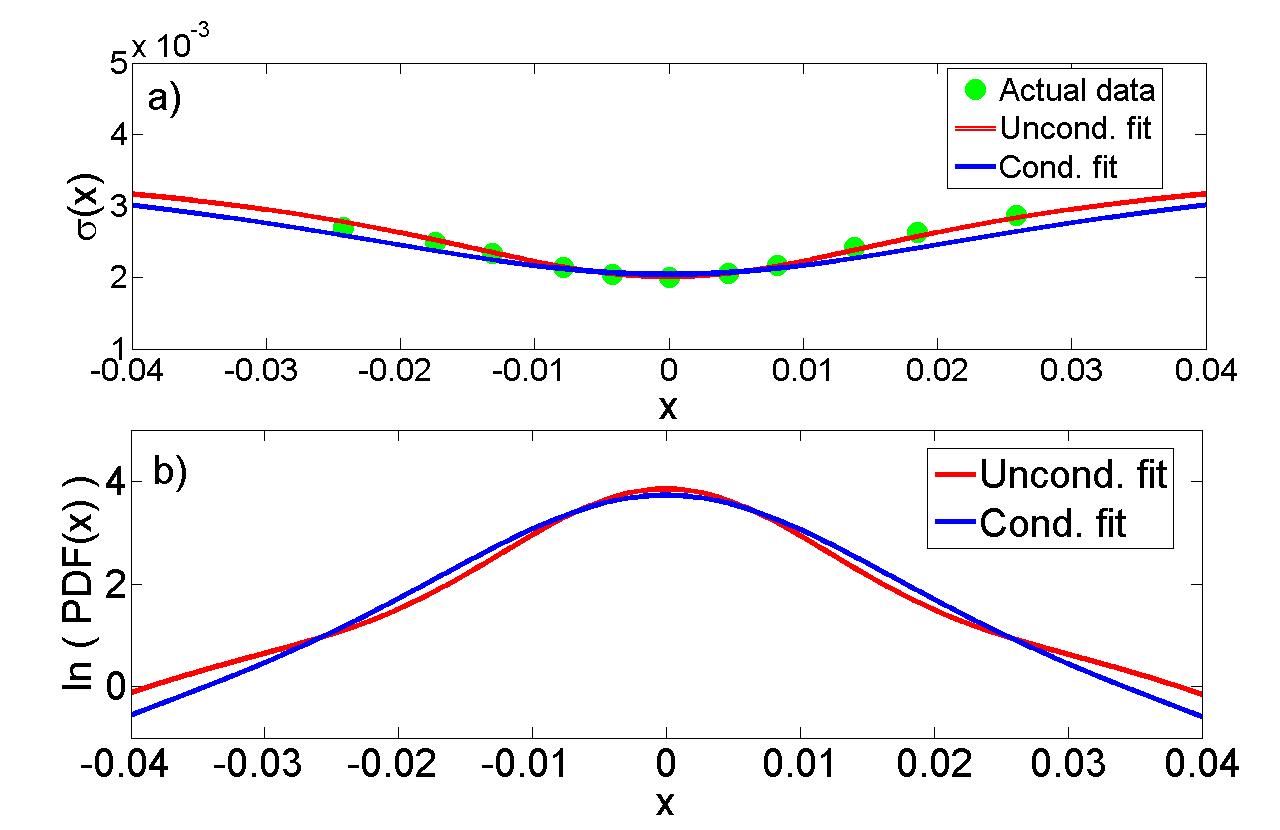}
\end{center}
\caption{We compare unconditional fit of the VS (a) 
and the implied PDF (b)  for a particular dataset 
(EURJPY, $T= 30$ days, downloaded on 21/10/2009  15:37) with
 the conditional one.  } 
\label{appl2} 
\end{figure}

\section{Conclusions}\label{conclusions}
We started from the pricing equation of the Black-Scholes model for an European call and we considered the effect of the VS correction on the implied PDF. Our approach comes from statistical physics and it is related to the adiabatic interpretation in~\cite{Spada}. We showed that similar fits of a VS could imply strong differences on the implied returns PDF with obvious consequences on the risk estimation. 
To obtain a stronger fitting procedure for the VS that can be compatible with the theoretical aspects of the model we first derived a relation between the exponential decay of the CCDF of returns and the parameters of the fitting function of the smile. Then, we exploit this relation to get a new fitting procedure that can be compatible with the historical data.
An interesting case is shown in Fig.~(\ref{appl}) where we compare the PDF of returns obtained by a standard fitting procedure of VS (unconditional fit) with the one obtained following the procedure described before (conditional fit). In this case the time to maturity is large, $T = 2520$, so we cannot get $\sigma^H$ and $\mu^H$ directly from the dataset but we extrapolate their values considering the relation $\mu \propto 1/\sqrt{T}$ and $\sigma \propto \sqrt{T}$. As evident, the unconditional fit generates an implied PDF with a relative minima never observed in actual data~\cite{Spada}, on the contrary the conditional fit generates a PDF more ``regular" that seems suitable for the description of actual PDF of returns. The price to pay in order to get a smooth PDF is related to the error for the smile fitting: the horizontal amplitude of the conditional fit is higher than the one required to get a suitable fit. This can be explained assuming that market makers overreact to extreme events when the time to maturity is large, estimating the volatility in a way that is not compatible with historical data. Besides, conditional fit is compatible with the skewness reduction claimed in~\cite{Ciliberti} to get a smile fitting more suitable to the historical data. \\
In conclusion we provide a new tool for the VS fitting that can be used to get a more coherent estimation of the parameters of fitting function, compatible with historical series and theoretical aspects of the model. A reliable estimate of the implied volatility has application in the risk management activities and in the pricing of exotic derivatives, where, in general, the
implied volatility is an input of more complex models.

\begin{figure}[t]
\begin{center}
\includegraphics[scale=0.2]{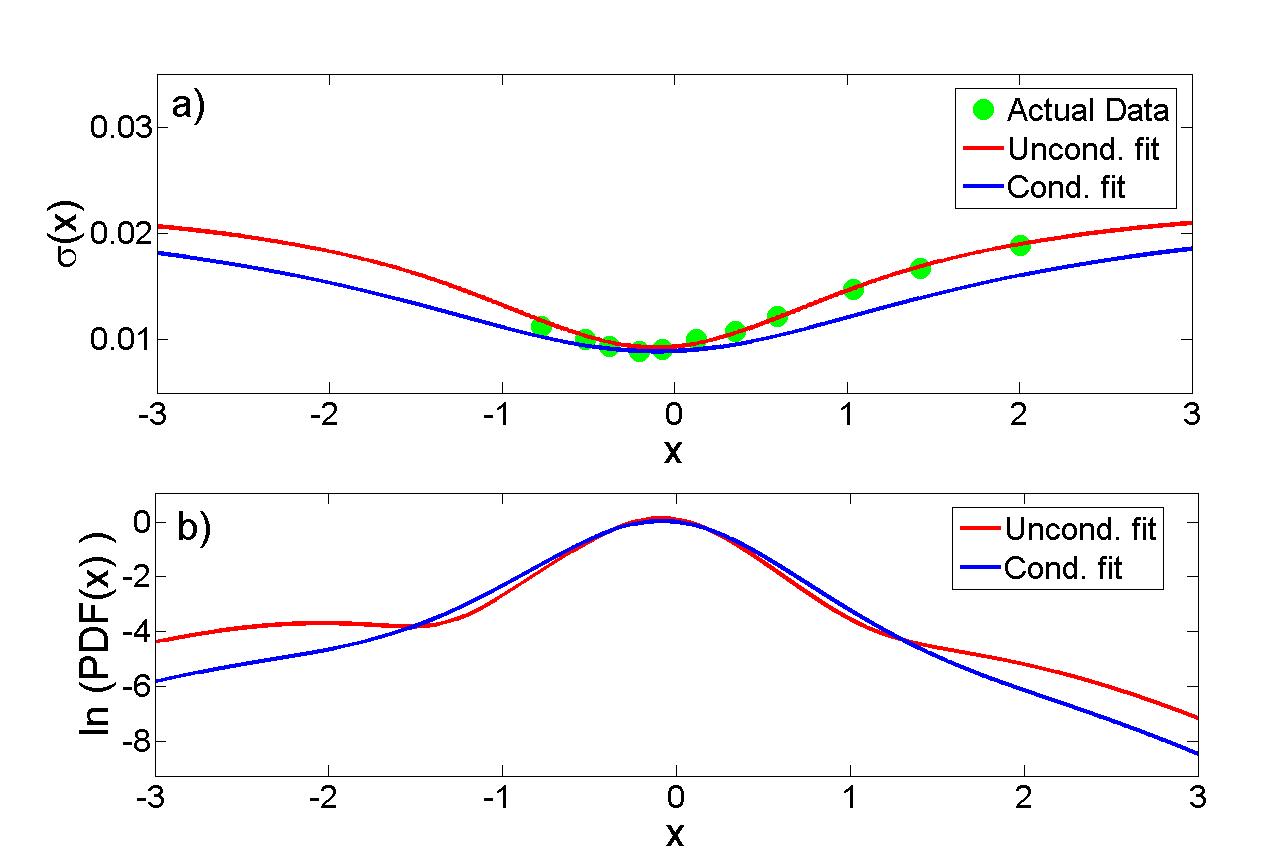}
\end{center}
\caption{We compare unconditional fit of the volatility smile (a) 
and the implied PDF (b)  for a particular dataset 
(EURJPY, $T= 2520$ days, downloaded on 21/10/2009  15:37) with the 
conditional one. To fix the historical decay parameters, 
$\mu_H, \sigma_H$, we find
 the values from actual data considering the relation 
$\mu \propto 1/\sqrt{T}$ (Eq.~\ref{mu_scaling}) and the standard
 scaling $\sigma \propto \sqrt{T}$. As evident, the 
PDF of conditional procedure does not present spurious minima and gives a distribution suitable to describe actual data. } 
\label{appl} 
\end{figure}

\section{Acknowledgments}
The work by GPB was carried out under the auspices of the NNSA of the U. S. DOE at LANL under Contract No. DEAC52-06NA2539.


\begin{thebibliography}{}
\bibitem{winer_note} The behavior of the Wiener process can be described on the finite time-interval, $\Delta t$, as $\Delta z  = \gamma \sqrt{\Delta t}$,where the random variable, $\gamma$, has the distribution function: $g(\gamma) = 1/\sqrt{2 \pi} \exp(-\gamma^2/2)$.
\end{thebibliography}
\end{document}